\documentclass[11pt]{article}
\usepackage[english]{babel}
\usepackage[utf8]{inputenc}

\usepackage{apacite}
\usepackage{authblk}

\usepackage{natbib}
\usepackage{graphicx, subcaption}

\title{A case for robust translation tolerance in humans and CNNs. A commentary on Han et al.}

\author[1]{Ryan Blything}
\author[1]{Valerio Biscione}
\author[1,*]{Jeffrey Bowers}

\affil[1]{University of Bristol, School of Psychological Science, Bristol, BS8 1TU, United Kingdom}
\affil[*]{j.bowers@bristol.ac.uk}

\date{}
\begin{document}

\maketitle


\citet{han2020scale} reported a behavioral experiment that assessed the extent to which the human visual system can identify novel images at unseen retinal locations (what the authors call “intrinsic translation invariance”) and developed a novel convolutional neural network model (an Eccentricity Dependent Network or ENN) to capture key aspects of the behavioral results.  Here we show that their analysis of behavioral data used inappropriate baseline conditions, leading them to underestimate intrinsic translation invariance. When the data are correctly interpreted they show near complete translation tolerance extending to 14$^{\circ}$ in some conditions, consistent with earlier work \citep{bowers2016visual} and more recent work \citep{Blything2020}.  We describe a simpler model that provides a better account of translation invariance.

\begin{figure}[!ht]
\centering
  \includegraphics[width=0.75\textwidth]{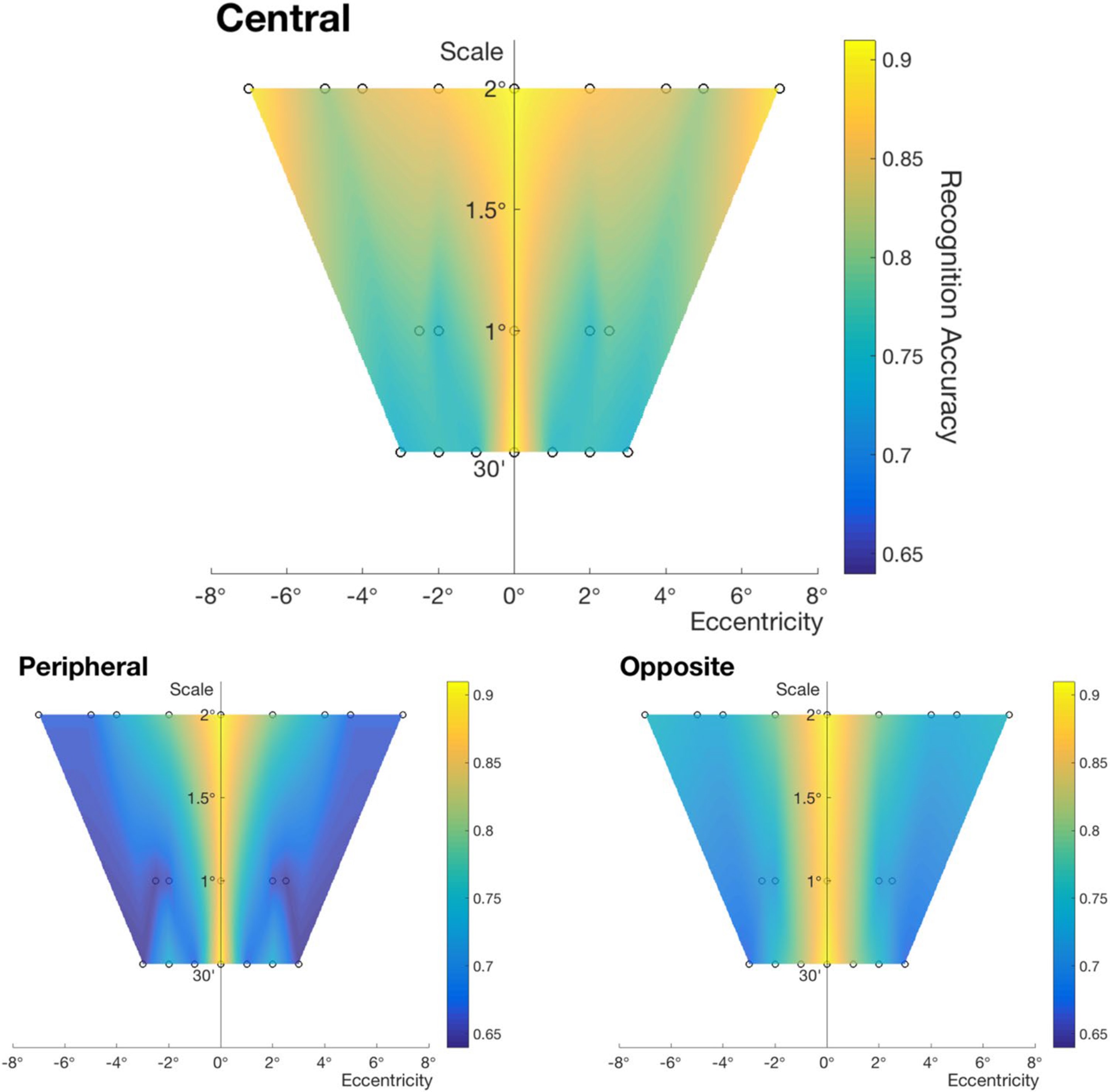}
\caption{\textit{Taken from \citet{han2020scale}.} Windows of invariance for different conditions. Same/different accuracy is shown in a color scale. The central window (top) depicts the results when target letters were presented at fixation and corresponding test letters in periphery at various scales and eccentricities. The peripheral window (bottom left) depicts the results when target letters were presented at various scales and eccentricities and corresponding test letters at fixation. The opposite window (bottom right) depicts the results when target and test letters are presented at various scales and eccentricities, with both target and test stimuli presented at the same distance from the center but in the opposite side of the visual field (or both stimuli presented at fixation at 0 degrees eccentricity). In all plots, the tested conditions are marked with circles and other data points are estimated using natural neighbor interpolation}
     \label{fig:Han_Beh}
   \end{figure}

\citeauthor{han2020scale} examined intrinsic translation invariance in humans by requiring non-Korean participants to classify Korean letters as same or different when a `target' letter was first flashed for 33ms followed by a `test' letter for 33ms two seconds later. Target and test letters were shown at fixation or at varying eccentricities, resulting in three translation conditions they labelled: \textit{Central} (central target, peripheral test), \textit{Peripheral} (peripheral target, central test), and \textit{Opposite} (peripheral target, test at opposite peripheral side). They also assessed the impact of image size (scale) on translation invariance, with letters subtending 30’, 1$^{\circ}$, or 2$^{\circ}$. In all cases, performance in these translation conditions was compared to a condition in which the target and test letters were presented at fixation. As illustrated in Figure \ref{fig:Han_Beh} (a reproduction of their Figure 3), they observed substantially reduced performance in the Peripheral and Opposite conditions for all scales, and reduced performance in the Central condition for the smaller stimuli. The authors took these findings to highlight substantial limitations of intrinsic translation invariance, especially for smaller stimuli. However, this conclusion is mistaken.  The reduced performance across the translation conditions was largely the product of poor visual acuity in peripheral vision.

The standard procedure for assessing translation invariance while controlling for visual acuity is to compare performance at a fixed eccentricity when target/test stimuli are presented at same retinal location to performance in the opposite visual fields (e.g., \citeauthor{afraz2008retinotopy}, \citeyear{afraz2008retinotopy}; \citeauthor{biederman1991evidence}, \citeyear{biederman1991evidence}; \citeauthor{cox2008does}, \citeyear{cox2008does}; \citeauthor{dill2001imperfect}, \citeyear{dill2001imperfect}; \citeauthor{dill1997role}, \citeyear{dill1997role}; \citeauthor{dill1998limited}, \citeyear{dill1998limited}).  In this way the target/test items have equal acuity, with a translation of 0$^{\circ}$ in the former condition, and two times the eccentricity in the later condition. However, this is not what is plotted in Figure \ref{fig:Han_Beh}. Rather, performance at 0$^{\circ}$ reflects performance when target and test stimuli are both presented at fixation where acuity is maximal, explaining the extremely high performance in the Central, Peripheral, and Opposite conditions at 0$^{\circ}$ for all image scales.  In all other translation conditions, the target or test letter (or both) were presented in peripheral vision, leading to a confound of translation with acuity.

Han et al. were aware of a possible confound with translation tolerance and visual acuity, but rejected acuity as the explanation for their findings based on their findings that Koreans participants (who were familiar with the letters) were highly accurate at most translation conditions. We disagree with this logic. Korean participants have learned to recognise degraded Korean letters in peripheral vision (e.g., when fixating on a central letter of a word the outer letters are projected in peripheral vision), and their success does not rule out acuity as the cause of the difficulty for non-Korean participants.  

\begin{figure}[!ht]
\centering
  \includegraphics[width=1\textwidth]{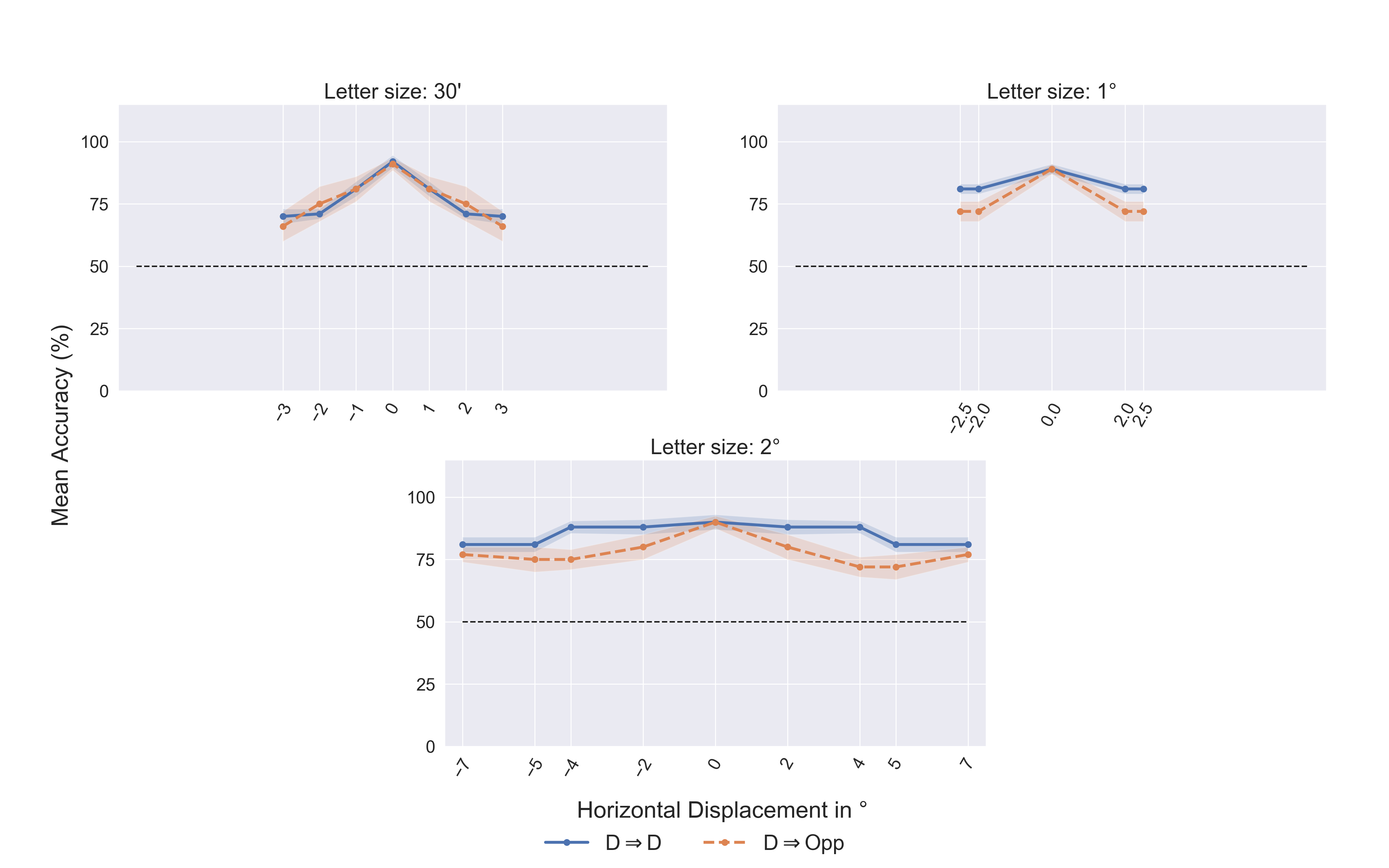}
\caption{Han Behavioral Data. Comparison of same location (D$\rightarrow$D) condition in blue vs. opposite location (D$\rightarrow$OPP) condition in yellow, replotted from their supplementary data.}. 
     \label{fig:Han_Beh_SI}
   \end{figure}

As noted above, critical data for assessing translation tolerance while controlling for visual acuity is to compare performance at a fixed eccentricity when target/test stimuli are presented at same or opposite visual fields. These data are reported in Han et al.'s Figure S1 (Supplementary Material), and are replotted here as Figure \ref{fig:Han_Beh_SI}.  In all cases performance was similar when the letters were flashed in the same (solid blue line) and opposite (dashed  yellow line) fields at a given eccentricity, with performance for the smaller stimuli reduced as a function of eccentricity. This shows that intrinsic translation invariance was near complete, with performance on the task largely limited by visual acuity.  The near complete translation invarience extending to 14 degrees for the largest Korean letters is consistent with \citep{bowers2016visual} who observed robust translation tolerance extending 13 degrees for images of unfamiliar images of 2D shapes that subtended 5 degrees, and \citep{Blything2020} who reported robust translation tolerance extending 18 degrees for images of unfamiliar 3D shapes that subtended 5 degrees.  


This reanalysis of the \cite{han2020scale} dataset undermines their Eccentricity Dependent Network (ENN) that was motivated to explain limited intrinsic translation invariance. Instead, the robust transation tolerance is more consistent with the standard CNN models that \cite{han2020scale} rejected, although those models still failed to capture the overall drop in acuity at peripheral locations. With this in mind, we trained a classic CNN network \cite[VGG16,][]{simonyan2014deep} on a dataset with injected pepper noise, probabilistically applied as a hyperbolic function of the distance to the center of the canvas.  The noise was designed to mimic the reduced acuity in peripheral vision
\citep{fovea} (Fig \ref{peppernoise}). Similarly to Han et al., we pretrained a network with a (noisy) MNIST handwritten dataset on different locations and scales (pretraining on translated objects is essential to obtain translation invariance on novel items, as shown in \citeauthor{Biscione2020}, \citeyear{Biscione2020}). We then re-trained this network on the set of Korean characters used by Han et al., divided in two groups of 15 characters each (same characters and grouping as in Figure 1 in their work, but we inverted the colours to keep consistency with the MNIST dataset). The network objective was to classify the characters from either of these groups. We used 4 different letter sizes: 10, 16, 22, and 28 pixels, on a 224x224 pixels canvas. Importantly, we trained the network by displaying the items at only one location, and tested the network on the same (D$\rightarrow$D) or opposite (D$\rightarrow$Opp) location (Fig. 3b-f). Notice that in the D$\rightarrow$Opp condition, the network is queried on a location in which it has never seen any Korean characters. By comparing D$\rightarrow$D to D$\rightarrow$Opp, we can infer the performance drop due uniquely to object displacement (translation tolerance). Results in Figure \ref{fig:PepperNet} show that intrinsic translation invariance is near perfect and bounded by visual acuity in the periphery.  The results succeed in qualitatively mimicking the relationship between object size, translation, and accuracy, found in humans.

One interesting finding from Han et al. that our CNN model does not explain is the behavioral asymmetry between 0$\rightarrow$D and D$\rightarrow$0. It is important to emphasize here that neither of these conditions assessed intrinsic translation invariance (performance was confounded with visual acuity), and the asymmetry may reflect properties of visual short-term memory rather than visual invariance.  For example, it may be more difficult to maintain a highly degraded image of an unfamiliar Korean letter in STM, and this selectively impaired performance in the same/different task when the first letter (the target) was presented in peripheral vision and had to be stored for 2 seconds. Previous behavioral studies that have avoided this confound with acuity have failed to observe this asymmetry (Bowers et al., 2016; Blything et al., in press).

In summary, \citet{han2020scale} have misinterpreted their behavioral data, and when intrinsic translation invariance is correctly assessed it is near complete for all size stimuli and bounded by visual acuity. These results are consistent with previous work (\citeauthor{bowers2016visual}, \citeyear{bowers2016visual}; \citeauthor{Blything2020}, \citeyear{Blything2020}), and are broadly consistent with standard CNNs when they are given inputs that capture human visual acuity. 

\newcommand{\ww}[0]{0.35}

\newcommand{\www}[0]{0.32}
  \begin{figure}
   \begin{subfigure}[t]{\www\textwidth}
      \includegraphics[width=\textwidth]{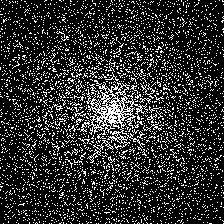}
      \caption{}
      \label{peppernoise}
    \end{subfigure}
    \hfill
    \begin{subfigure}[t]{\www\textwidth}
      \includegraphics[width=\textwidth]{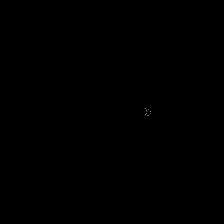}
      \caption{}
    \end{subfigure}
    \hfill
    \begin{subfigure}[t]{\www\textwidth}
      \includegraphics[width=\textwidth]{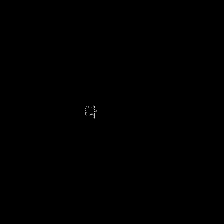}
      \caption{}
    \end{subfigure}
    \hfill
    \begin{subfigure}[t]{\www\textwidth}
      \includegraphics[width=\textwidth]{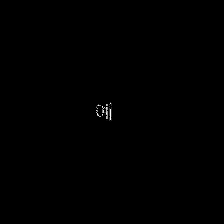}
      \caption{}
    \end{subfigure}
    \hfill
    \begin{subfigure}[t]{\www\textwidth}
      \includegraphics[width=\textwidth]{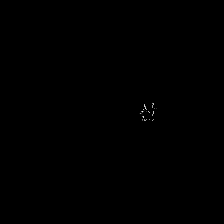}
      \caption{}
    \end{subfigure}
    \hfill
    \begin{subfigure}[t]{\www\textwidth}
      \includegraphics[width=\textwidth]{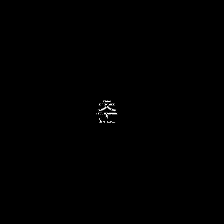}
      \caption{}
    \end{subfigure}
    \hfill
   
    \caption{(a) An example of randomized pepper noise used in our experiment, here applied on a white canvas. This is the same type of noise applied to all images in our dataset.  Notice how the amount of noise increases the farther away from the center. (b-f) Examples of input images for the network, at different spatial locations and different sizes. (b-c) are objects of size 10 and 16 pixels, slightly displaced from the center (b is on the right, c on the left of the canvas center). Due to their small sizes, the stimulus are barely visible even with limited translation. (d-f) are example of objects with larger sizes, more visible even when translated.}
  \end{figure}

\begin{figure}[h]
\centering
  \includegraphics[width=0.95\textwidth]{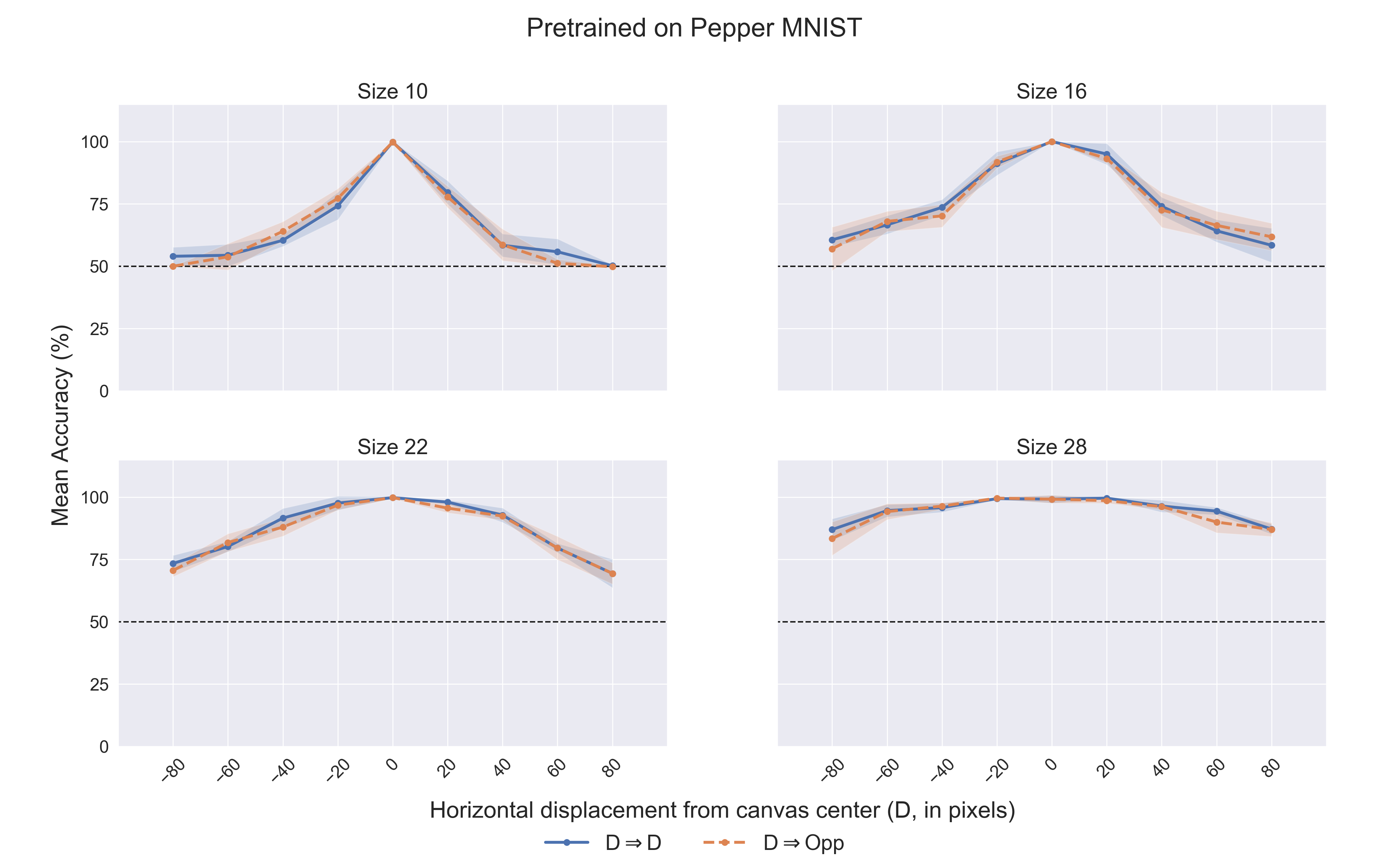}
\caption{\t Results from the modelling experiment presented in this work. A VGG16 network is trained on peripherally degraded images along horizontal locations of the canvas. For each trained location, the network is then tested on the same location (D$\rightarrow$D) or on the opposite location (D$\rightarrow$Opp). When accounting for limitation due to image degradation, we can observe that the network is highly tolerant to translation to almost the whole canvas.}
     \label{fig:PepperNet}
   \end{figure}

\clearpage

\bibliographystyle{apacite}

\end{document}